\documentclass[twocolumn, prb, superscriptaddress,
  showpacs]{revtex4-1} 
\usepackage{amsmath, array} 
\usepackage{graphicx}
\usepackage{grffile}
\usepackage{hyperref} 
\usepackage{color}

\newcommand{\EqLabel}[1]{\label{#1}}

\begin{document}
\title{Role of long-range  coupling on the properties of single polarons in models with dual electron-phonon couplings }

\author{Monodeep Chakraborty} \affiliation{Centre for Quantum Science and Technology, Chennai Institute of Technology, Chennai, India-600037}
\author{Sankeerth S. Narayan} \affiliation{Computer Science Engineering Department, Chennai Institute of Technology, Chennai, India-600037}
\author{Vigneshwaran R.} \affiliation{Mechatronics Engineering Department, Chennai Institute of Technology, Chennai, India-600037}
\author{Mona Berciu}
\affiliation{Department of Physics and Astronomy, University
  of  British Columbia, Vancouver, British Columbia, Canada,
  V6T 1Z1} \affiliation{Quantum Matter Institute, University
  of British Columbia, Vancouver, British Columbia, Canada,
  V6T 1Z4} 

\begin{abstract}
  We use the Variational Exact Diagonalization to investigate the single polaron properties for four different dual models, combining a short-range off-diagonal (Peierls) plus a longer-range diagonal (Holstein or breathing-mode) coupling. This allows us to investigate the sensitivity of various polaron properties both to the range of the diagonal coupling, and to the specific diagonal coupling chosen. We find strong sensitivity to the range for all duals models as the adiabatic limit is approached, however considerable sensitivity is observed for some quantities even for large phonon frequencies. Also, strong dependence of the results on the specific form of the diagonal coupling is observed everywhere in the parameter space. Taken together, these results suggest that a careful consideration must be given to the specific coupling and its proper range, when quantitative comparisons with experiments are sought.
\end{abstract}
\date{\today}


\maketitle

\section{Introduction}

The consequences of the interplay between bosonic and fermionic degrees of freedom are of interest in most areas of physics. In solid state physics, polarons are one of the earliest and most quintessential examples of such physics, where the coupling between electronic and phononic degrees of freedom results in the formation of a quasiparticle (the polaron) comprising a charge carrier dressed by a cloud of phonons. Physically, this dressing describes the lattice distortion induced by the charge carrier.

One limit of this problem where there has been substantial progress is on understanding the properties of single polarons (at vanishing charge carrier concentration)  for a variety of electron-phonon (e-ph) couplings. While most of the earlier work was dedicated to continuum models with long-range el-ph coupling like the Fr\"ohlich model,\cite{Frohlich} the advent of considerable computational power together with the development of numerical methods in the last few decades has made the study of lattice models feasible. The most well studied lattice model with electron-phonon coupling is the Holstein model, which assumes local (on-site) coupling between the density of carriers and the phonon displacement operator.\cite{Holstein1,Holstein2}

For some time, it was assumed that the properties of Holstein single polarons are qualitatively representative of the properties all  single polarons, irrespective of the details of the e-ph coupling. More recently, however, it has become clear that this is not the case. There are, in fact, two different classes of electron-phonon couplings, differentiated based on their physical origin. The Holstein model is part of the class of so-called $g(q)$ models, where the coupling arises from the modulation of the on-site energy of the electron due to the motion of nearby ions. Because the underlying interaction is of (screened) Coulomb nature, this coupling is proportional to the density of electrons, {\em ie} it is diagonal in electron operators in real space. This is the reason why upon Fourier transforming, its vertex $g(q)$ cannot depend on the electron momentum $k$, only on the momentum $q$ of the emitted/absorbed phonon. To date, it appears that $g(q)$ models are indeed qualitatively similar to one another insofar as the properties of their single polarons are concerned. In particular, they all show a smooth crossover between the weak- and strong-coupling limits,\cite{GL} and the polaron effective mass increases monotonically as the coupling strength is increased.\cite{Holstein3}

The second class of models, the so-called $g(k,q)$ models, arise from modulation of the hopping integrals of the electron due to the motion of nearby ions; here, dependence of the electron momentum $k$ appears from the fact that the electronic part is off-diagonal in real space.
The most well-known such model is the Barisi\'c-Labb\'e-Friedel model,\cite{SSH1,SSH2,SSH3} also known as the Su-Schrieffer-Heeger\cite{SSH4,SSH5} (in the context of polyacetylene) or more generally, the Peierls model. For brevity, we will refer to this as the Peierls coupling from now on. In Ref. \onlinecite{Dominic}, it was shown that the Peierls single polaron has  properties that are qualitatively different from those of the $g(q)$ polarons and this was later confirmed for a variety of related models.\cite{EdM,BP1,BP2} In particular, in some $g(k,q)$ models including the Peierls coupling, sharp transitions of the polaron  ground-state between weak- and strong-coupling limits have been identified,\cite{Lieb1,Lieb2,Lieb3}  and it was also found that single polarons can remain very light at strong couplings.\cite{Dominic,BP1,BP2}

The question of what happens upon combining a $g(q)$ and a $g(k,q)$ coupling has also been explored, although much less. It is clear that depending on the symmetries of the two couplings, non-trivial ``interference'' effects are possible. For example, a combination of the $g(q)$  breathing-model (BM) coupling\cite{BBM1,BBM2}  and the Peierls coupling has revealed the appearance of two sharp transitions in the single polaron ground-state properties\cite{Roman}, even though the BM single polaron cannot have a transition,\cite{GL} while the Peierls single polaron only has one.\cite{Dominic} On the other hand, a combination of Holstein and SSH couplings was found to be more trivial, in the sense that the results are essentially the sum of the results for the individual couplings,  there is no 'interference' leading to new physics.\cite{DominicPRB}

All this work, however, has focussed on short-range el-ph couplings, where the electron operators are coupled to phonons operators either on the same site, or a nearest-neighbor site. To our knowledge, there has been no study of a dual model where the $g(q)$ coupling has a long-range in real space. We note that the extended Holstein model (EHM) was studied on its own to contrast the results of this long-range lattice model with those of the continuum Fr\"ohlich model,{\cite{Feshke,M1} but not in combination with a $g(k,q)$ coupling. We also note that a long-range $g(k,q)$ coupling is less physical, which is why we do not consider it (this issue is discussed in more detail below,  where the specific models are introduced).

Studies of longer-range el-ph couplings have become very timely given recent work that supports the idea that the longer-range nature is essential for quantitative modeling of one-dimensional (1D) cuprate chains.\cite{ZX,Yao} This motivates us to study in this work the properties of single polarons in several 1D dual models whose $g(q)$ part has a variable range, so as to understand how/when is the longer range part relevant, and how sensitive the results are to other details of the specific $g(q)$  coupling chosen. Of course, these single polaron results are relevant in the vanishing carrier limit, whereas the cuprate chains mentioned above are at a finite concentration. Our results are therefore not directly relevant for those systems. Nevertheless, we believe that they highlight important conclusions about the modeling of such systems at any charge carrier concentration, in addition to increasing the general knowledge of single polaron physics.

Our work is organized as follows: in Section II we introduce the various models we study and explain their underlying assumptions. In Section III we briefly review the well-established Variational Exact Diagonalization (VED) method we used to study these dual models. Section IV contains our representative results, while Summary and Conclusions are formulated in Section V.

\section{Models}

All the models investigated in this work are for an infinite  chain of identical atoms, with lattice constant $a=1$. In all cases, the bare Hamiltonian is assumed to be:
\begin{equation}
\EqLabel{H0}
{\cal H}_0={\cal T} + {\cal H}_{\rm ph}
\end{equation}
where
\begin{equation}
\EqLabel{2}
{\cal T} =-t \sum_{i}^{}\left( c_i^\dagger c_{i+1} + h.c.\right)
\end{equation}
%%%%%%%%%%%%%%%%%%%%%%%%%%%%%%%%%%%%%%%%%%%%%%%%%%%%%%%%%%%%%%%%%%%%%%
describes nearest-neighbor (nn) hopping of
the electron, and $c_i$ annihilates an electron at
site $i$ (we ignore the spin projection because it is irrelevant in single polaron physics in the absence of spin-orbit coupling). Correlations are irrelevant because there is a single carrier in the system.
The phonons, with creation operators
$b_i^\dagger$, are described as an Einstein mode of frequency
$\Omega$ (we set $\hbar=1$ throughout):
%%%%%%%%%%%%%%%%%%%%%%%%%%%%%% EQUATION %%%%%%%%%%%%%%%%%%%%%%%%%%%%%%
\begin{equation}
\EqLabel{3}
{\cal H}_{\rm ph} = \Omega \sum_{i}^{} b_{i}^\dagger b_i.
\end{equation}
Various forms of the el-ph coupling are discussed next.

\subsection{The extended Holstein coupling}

We begin with a brief review of the extended Holstein model (EHM) $g(q)$ coupling both to establish a link to existing previous work,{\cite{Feshke,M1} and also because a similar model was used in  Refs. \onlinecite{ZX,Yao} for  quantitative modeling of el-ph coupling in 1D CuO chains.

The extended Holstein coupling has the standard form \cite{Feshke,M1}:
%%%%%%%%%%%%%%%%%%%%%%%%%%%%%% EQUATION %%%%%%%%%%%%%%%%%%%%%%%%%%%%%%
\begin{equation}
\EqLabel{4}
 \hat{V}_{eh}=g\Omega \sum_{ij}f_{j} c_i^\dagger c_i (b_{i+j}^{\dag}
+ b_{i+j})
\end{equation}
%%%%%%%%%%%%%%%%%%%%%%%%%%%%%%%%%%%%%%%%%%%%%%%%%%%%%%%%%%%%%%%%%%%%%%^M
where $g$ is a dimensionless parameter quantifying the overall strength of the EH coupling, while the spatial dependence is described by:
\begin{equation}
f_{j} = \frac{\Theta(M-|j|)}{(j^{2} +1 )^{\frac{3}{2}}}.
\end{equation}
The Heaviside function allows us to control the range of this model depending on the cutoff $M$. If $M$ is chosen so that the coupling is finite only for $j=0$, we regain the standard Holstein model (on-site coupling). Increasing $M$ allows us to study increasingly longer-range coupling, to see how this influences the results. In the following, we characterize the coupling range with (i) the label $1$ for the Holstein model ($M=0.5$); (ii) the label 3 when the extended coupling includes the three sites $j=0, \pm 1$ ($M=1.5$); (iii) the label 5 when the extended coupling includes the five sites $j=0, \pm 1, \pm 2$ ($M=2.5$); and (iv) the label 7 when the extended coupling includes the seven sites $j=0, \pm 1,\pm 2, \pm 3$ ($M=3.5$). 

\begin{figure}[t]
\centering \includegraphics[width=0.7\columnwidth,angle=-90]{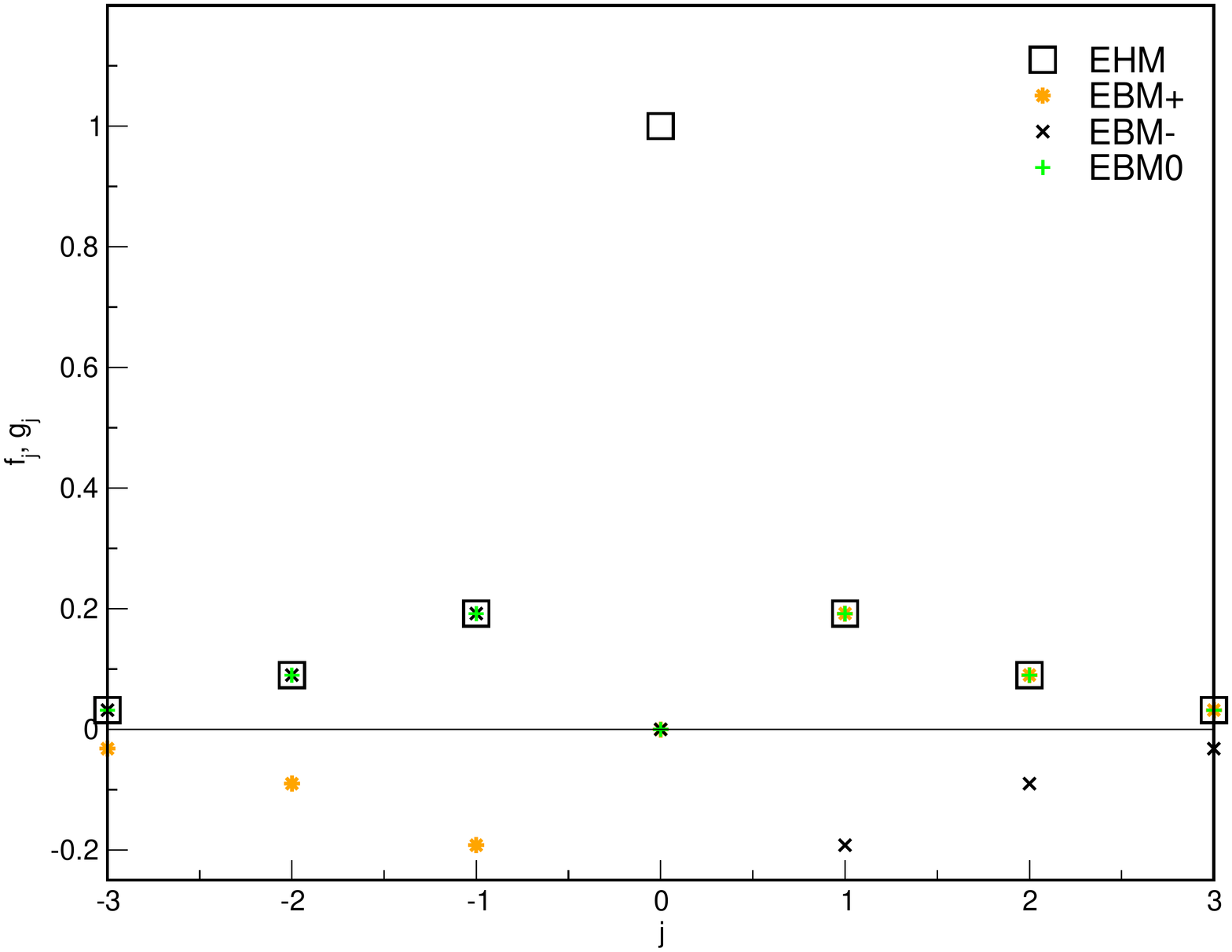}
  \caption{Plots of $f_{j}$ for the EHM, and of $g_j$  for the three EBM models considered in this work.
  \label{fig1}}
\end{figure}

To make meaningful comparisons for these different ranges, we need to rescale the magnitude $g$ as a function of the cutoff $M$, so that the physical energy scale -- the polaron formation energy in the single-site limit -- remains the same. For EHM, the polaron formation energy (at $t=0$) for a given cutoff $M$ is
\begin{equation}
{\epsilon}_{p} = - g^2 \Omega \sum_{|j|<M }f_{j}^{2}
\end{equation}
Comparing this energy to the ground-state energy $-2t$ of the bare electron defines the  dimensionless electron-phonon coupling $\lambda=-{\epsilon_p}/{2t}$. For the Holstein model, then, $\lambda_h=g^2/(2t\Omega)$. In order to have the same overall $\lambda_{eh} = \lambda_h$ for different coupling ranges in the EHM, we rescale the value of $g$ such that
\begin{equation}
  \label{gHol}
  g=\sqrt{\frac{2t\lambda_h}{\Omega\sum_{j}f_{|j|<M}^{2}}}.
  \end{equation}

\subsection{The Peierls coupling}

The $g(k,q)$ Peierls coupling is due to linear modulations of the nn hopping because of displacements of the atoms involved in the hopping:
\begin{equation}
\EqLabel{4}
 \hat{V}_{p}=\alpha\sum_{i}^{}\left( c_i^\dagger c_{i+1} + h.c.\right)\left( b_{i+1}^\dagger+b_{i+1} - b_{i}^\dagger-b_{i}\right).
\end{equation}
%%%%%%%%%%%%%%%%%%%%%%%%%%%%%%%%%%%%%%%%%%%%%%%%%%%%%%%%%%%%%%%%%%%%%%
Following previous work,\cite{Dominic} we  use the dimensionless effective coupling $\lambda_{p} = 2{\alpha}^2/(\Omega t)$ to characterize the strength of the Peierls coupling. 

For the Peierls coupling we do not consider a longer-range extension. That can be accomodated by VED, however it is not very physical given that such terms would arise from modulations of the longer-range hopping integrals. The hopping integrals are assumed to decrease exponentially with the distance between the sites, which is a much faster decrease than the power-law screening of the Coulomb interactions responsible for the $g(q)$ couplings. This is why we believe that the primary source of long-range coupling must be coming from the $g(q)$ model, and we only consider this option in our work.

\subsection{The EHM+P dual model}

The first dual model that we study combines the extended Holstein+Peierls couplings. Its Hamiltonian is:
%%%%%%%%%%%%%%%%%%%%%%%%%%%%%% EQUATION %%%%%%%%%%%%%%%%%%%%%%%%%%%%%%
\begin{equation}
\EqLabel{EHP}
{\cal H}_{eh+p}={\cal H}_0 + \hat{V}_{p} + \hat{V}_{eh}
\end{equation}
%%%%%%%%%%%%%%%%%%%%%%%%%%%%%%%%%%%%%%%%%%%%%%%%%%%%%%%%%%%%%%%%%%%%%%
and the strenghts of the couplings will be characterized by a $\lambda_p$ and a $\lambda_{eh}$, while the EHM range is set by $M$ as discussed above.
We note that this extends the work of Ref. \onlinecite{DominicPRB} where the combination of Peierls and Holstein models was investigated, providing another point of contact to previous work, and another way to validate our results. The comparison with those results will illustrate the effects of longer-range EHM coupling on the properties of the resulting polaron. On the other hand, contrasting these results with those of the EHM coupling will also allow us to understand the effects of adding  Peierls coupling in the model.

However, although the EHM model was already suggested to be relevant for 1D chains while the Peierls coupling is generically expected to appear in any system with el-ph coupling, their combination is not very physical. This is because Holstein-like couplings are due to an internal distortion of the 'polar molecule' assumed to be located at each lattice site, when an additional electron visits it. As such, this coupling is to the 'vibron' describing the internal distortion of the molecule, whereas the Peierls coupling is to actual phonons describing the displacements of sites from their equilibrium position.

To mitigate for this issue, we propose another $g(q)$ model which does couple to the same actual phonons like the Peierls model, so that their combination is more physical. For completeness, we also note that for a chain with a strictly one-site basis, there is a single longitudinal acoustic phonon mode, there is no optical mode that could be modelled as an Einstein phonon, as we do here. Nevertheless, we  continue to use the simplified Einstein phonon mode instead of a more realistic acoustic phonon mode. Physically, this is because both the Peierls coupling and the extended breathing-mode coupling described below vanish for $q=0$, i.e. there is no coupling to the gapless phonons at the center of the Brillouin zone. In contrast, the coupling is strong to the  phonons at the Brillouin-zone boundary at $q=\pi$, which describe anti-phase motion of consecutive ions. In other words, the phonons to whom the coupling is strong are similar to optical phonons (gapped and describing anti-phase motion). The second reason to use an Einstein model is because this is what was used in all the previous work we contrast our results with. By using the same phonon model we insure that any differences in results are due to the changes in the coupling. 

\subsection{The extended breathing-mode coupling}

In this model, the electron-phonon coupling is due to the modulation of the screened Coulomb interaction $V(i-j)$ between the electron at site $i$ and the ion at site $j$, because of the motion of both sites. To linear order, this leads to the extended breathing-mode (EBM) coupling:
$$
\hat{V}_{ebm}= g_{bm} \sum_{i,j} g_{j-i} c_i^\dagger c_i  (b_{j}^{\dag}
+ b_{j}-b_{i}^{\dag} - b_{i} )
$$
where $g_{j-i} = - g_{i-j}$ is  an odd function. As a result, the total  coupling to the phonons at the carrier site $j=i$ vanishes, and the EBM coupling can be written in a form similar to the EHM coupling:
\begin{equation}
\EqLabel{4}
 \hat{V}_{ebm}= g_{bm} \sum_{i,j} g_{j} c_i^\dagger c_i (b_{i+j}^{\dag}
+ b_{i+j})
\end{equation}
however now $g(j)$ is an odd function (as opposed to an even one, for the EHM), so  $g_0=0$. So while in the EHM  the on-site coupling is the largest, for the EBM  the on-site coupling vanishes. 

To facilitate comparisons between results of the EBM and the EHM, in the following we take:
\begin{equation}
  g_{j}=  ( 1- \delta_{j,0}) \chi_j f_j
  \label{gj}
\end{equation}
so that the long-range decay is similar. The coefficients $\chi_j$ are just signs. Specifically, we study three possible choices: (i) $\chi_j = j/ |j|$, describing a case where there is overall attraction between the electron and the other ions in the lattice (we always take $g_{bm}>0$). We call this the 'EBM+' case; (ii) $\chi_j = - j/ |j|$, describing a case where there is overall repulsion between the electron and the other ions in the lattice. We call this the 'EBM-' case; and (iii) $\chi_j=1$. This last choice is {\it not} leading to an odd $g_j$ function so it is not a proper breathing-mode coupling, instead it describes the EHM coupling without the on-site term. We will study it for comparison purposes, and we label it (somewhat improperly) 'EBM0'. Like for the EHM, we use the labels 1, 3, 5 or 7 to characterize the  range of the extended coupling.

To characterize the strength of the EBM coupling, we proceed as follows. First, we note that we could use a formula similar to the one used for the $EBM$ coupling. However, that choice would mean that if $\lambda_{eh}=\lambda_{ebm}$, then $g_{bm}\gg g$ in order to compensate for the absence of the on-site coupling. This is not suitable for our purpose, which is to compare models with equal magnitudes of the longer range coupling, to see how important are the details of the chosen form.

This is why we will characterize $g_{bm}= g$ by saying (rather improperly) that $\lambda_{eh}=\lambda_{ebm}$. In other words, the magnitude $g_{bm}$ for a given value of $\lambda_{ebm}$ is taken to be equal to the magnitude $g$ of the EHM of equal range, when $\lambda_{eh}=\lambda_{ebm}$. Comparing EHM with EBM0 will allow us to gauge the importance of the on-site coupling (present for EHM and absent for EBM0) for equally strong longer-range coupling, while comparing EBM0 with EBM$\pm$ will reveal the importance of the even vs. odd coupling under inversion.

\subsection{The EBM+P dual model}
The second, more physical dual model that we investigate therefore combines the EBM+P couplings:
%%%%%%%%%%%%%%%%%%%%%%%%%%%%%% EQUATION %%%%%%%%%%%%%%%%%%%%%%%%%%%%%%
\begin{equation}
\EqLabel{EBM}
{\cal H}_{ebm+p}={\cal H}_0 + \hat{V}_{p} + \hat{V}_{ebm}
\end{equation}
%%%%%%%%%%%%%%%%%%%%%%%%%%%%%%%%%%%%%%%%%%%%%%%%%%%%%%%%%%%%%%%%%%%%%%
As mentioned just above, there are actually 3 distinct variants of the EBM depending on the choice of $\chi_j$. All three are investigated below.

\section{Variational Exact Diagonalization}
The variational approach based on exact diagonalization (VED) is one of the most successful numerical methods for studying electron-phonon problems in different paradigms, especially in the dilute regime (with one or few electrons on an infinite lattice).

The VED is well established {\cite{BKT,BT,BTB} so we provide here only a very brief summary. We start 
from a one-electron Bloch state with a given momentum $k$ in the phonon vacuum of an infinite  chain, $|k \rangle \propto \sum_n \exp(i k n) c^\dagger_n|0\rangle$. Additional basis states are generated by repeated 
action of the off-diagonal pieces of the Hamiltonian on this initial state. If a new configuration (describing a new distribution of phonons relative to the electron) is generated, only one copy of is retained because translational symmetry is automatically taken care off. One of the biggest strengths of this method is that it allows for computation of polaron properties 
at any $|k|\le \pi $ in the  Brillouin zone,  rather than being limited to multiples of $\frac{2\pi}{N}$, which would be the case for doing 'traditional' ED on a finite $N$-sites chain.

The largest variational basis that we have used in this study has $N_h$=$14$ (i.e., the Hamiltonian was applied $14$ times on the initial state and all new configurations thus generated were retained) for the coupling involving onsite interaction (label '1' ), resulting in a basis with $N$=$15646560$ configurations; $N_h=13$ for coupling involving up to  the nearest-neighbor and the 2nd nearest-neighbor sites (labels '3' and  '5'), resulting in bases with $N$=$6233884$ and  $N$=$13770156$, respectively; and $N_h=12$ for coupling involving up to 3nd nearest-neighbor sites (label '7'), resulting in a basis with $N$=$9826209$ configurations. The numerical convergence of the results we present was verified by comparing these results obtained with the largest base, against those from the previous step (for example, $N_h=14$ vs $N_h=13$ for the short-range models).

One of the biggest strengths of the VED method is its accuracy in
interemediate coupling regime where peturbative methods are very
problematic. There is a constraint on the size of $N_h$ that can be implemented for any given
microscopic Hamiltonian, and hence the numerical accuracy becomes an
issue in the very strong coupling regime. For the couplings used in this work we have achieved convergence using the traditional VED. We note that methods like
  Lang-Firsov VED \cite{Alt2,M1} and incorporation of shifted
  oscillator states in conjunction with VED \cite{Alt1} have been successfully used to deal with the Holstein Hamiltonian at strong couplings. Similarly, the method of SC-VED \cite{M2,M3,M4} has
  been shown to be quite successful in numerical challenging
  parametric regimes for Holstein, extended Holstein as well as Edwards
  model. These extensions could be used to extend the study of the dual models proposed here, to stronger couplings than we consider.

\begin{figure}[t]
  \centering \includegraphics[width=0.7\columnwidth,angle=-90]{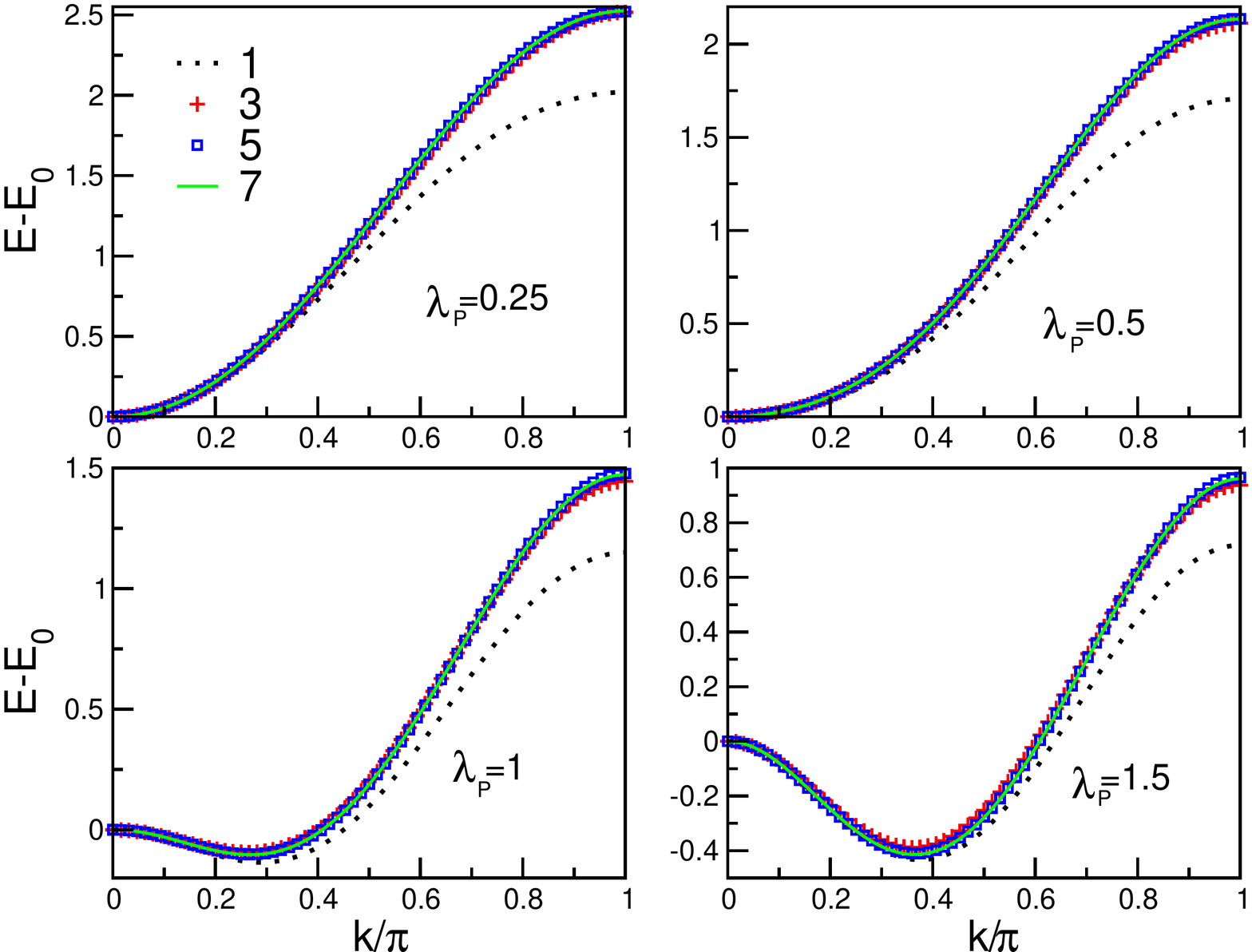}
  \caption{Shifted polaron dispersion $E_k-E_0$ vs $k$ at $\Omega$=$3$, $\lambda_{eh}$=$0.5$ for different ranges of the extended Holstein coupling: 1 (black  dotted line), 3 (red $+$ symbols), 5 (blue open squares), 7 (green solid lines). The four panels correspond to four different values of $\lambda_{P}$.
  \label{fig2}}
\end{figure}

\section{Results}

\subsection{EHM+P results}

First, we analyze the properties of polarons in the dual coupling model combining the extended Holstein + Peierls couplings. We set the energy scale to be $t=1$. We generated results for $\Omega= 0.5, 1$ and $3$, covering the crossover from weakly adiabatic to weakly anti-adiabatic regime. Figures \ref{fig2} and \ref{fig4} show representative results for a variety of coupling strengths and ranges, for the largest and smallest $\Omega$ values considered.

In all of these panels, the black dotted curves (label 1) are for the (on-site) Holstein+Peierls dual models, and are in great agreement with those generated in Ref. \onlinecite{DominicPRB} using  Bold Diagramatic Monte Carlo and the variational Momentum Average approximation. In particular, they show the expected change in the dispersion from one with a ground-state (GS) at $k_{gs}=0$ at weak Peierls coupling into one with a doubly degenerate GS with a $\pm k_{gs}\ne 0$ at strong Peierls coupling. The reason for this change is the dynamical generation of 2nd nearest-neighbor, phonon-mediated, hoppings in the presence of Peierls coupling. \cite{Dominic} 

\begin{figure}[t]
  \centering \includegraphics[width=0.7\columnwidth,angle=-90]{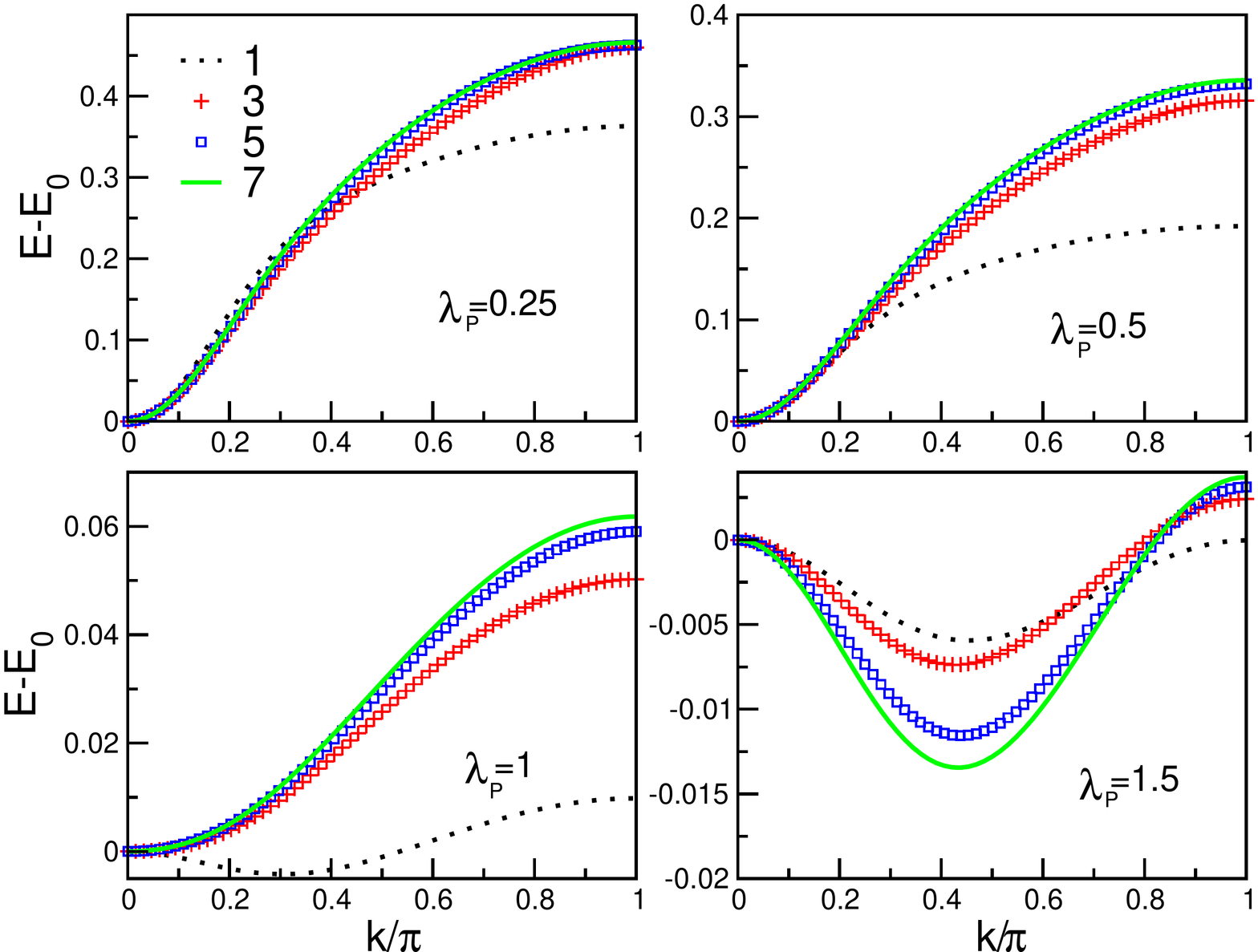}
  \caption{Same as in Figure \ref{fig2} but for $\Omega=0.5$.
  \label{fig4}}
\end{figure}

As the range of the extended Holstein coupling is increased (labels 3, 5, 7 for couplings including nn, 2nd nn, and 3rd nn neighbors respectively) such that  $\lambda_{eh}=0.5$ remains fixed, Fig. \ref{fig2} shows a non-trivial quantitative change with the addition of the extended  coupling to the nn sites, but then results are converged and no longer change if the range is further increased. By contrast, for the smaller $\Omega=0.5$ in Fig. \ref{fig4}, we see that dispersion continues to change as the range is increased. For weak Peierls coupling   $\lambda_P\lesssim 0.5$, convergence is achieved at the 2nd nn range, but for the stronger Peierls coupling the dispersion is not yet converged at 3rd nn range. Furthermore, the quantitative changes are much more substantial than was the case for the  results of Fig. \ref{fig2}.

At first sight, this strong sensitivity to the range of the extended coupling is rather unexpected, considering how relatively weak is the coupling to the 2nd and especially the 3rd nn sites, compared to the on-site one (see Fig. \ref{fig1}). However, it is known that the spatial size of the polaron cloud increases as $\Omega$ decreases, especially for moderate Holstein coupling such as the one used here, $\lambda_h=0.5$. One expects the structure of a  more extended cloud to be quite sensitive to the details of the extended coupling, as the probabilities for various configurations could be reshuffled significantly for an extended vs. a local coupling. This appears to be consistent with the results reported so far.

\begin{figure}[t]
  \centering \includegraphics[width=0.7\columnwidth,angle=-90]{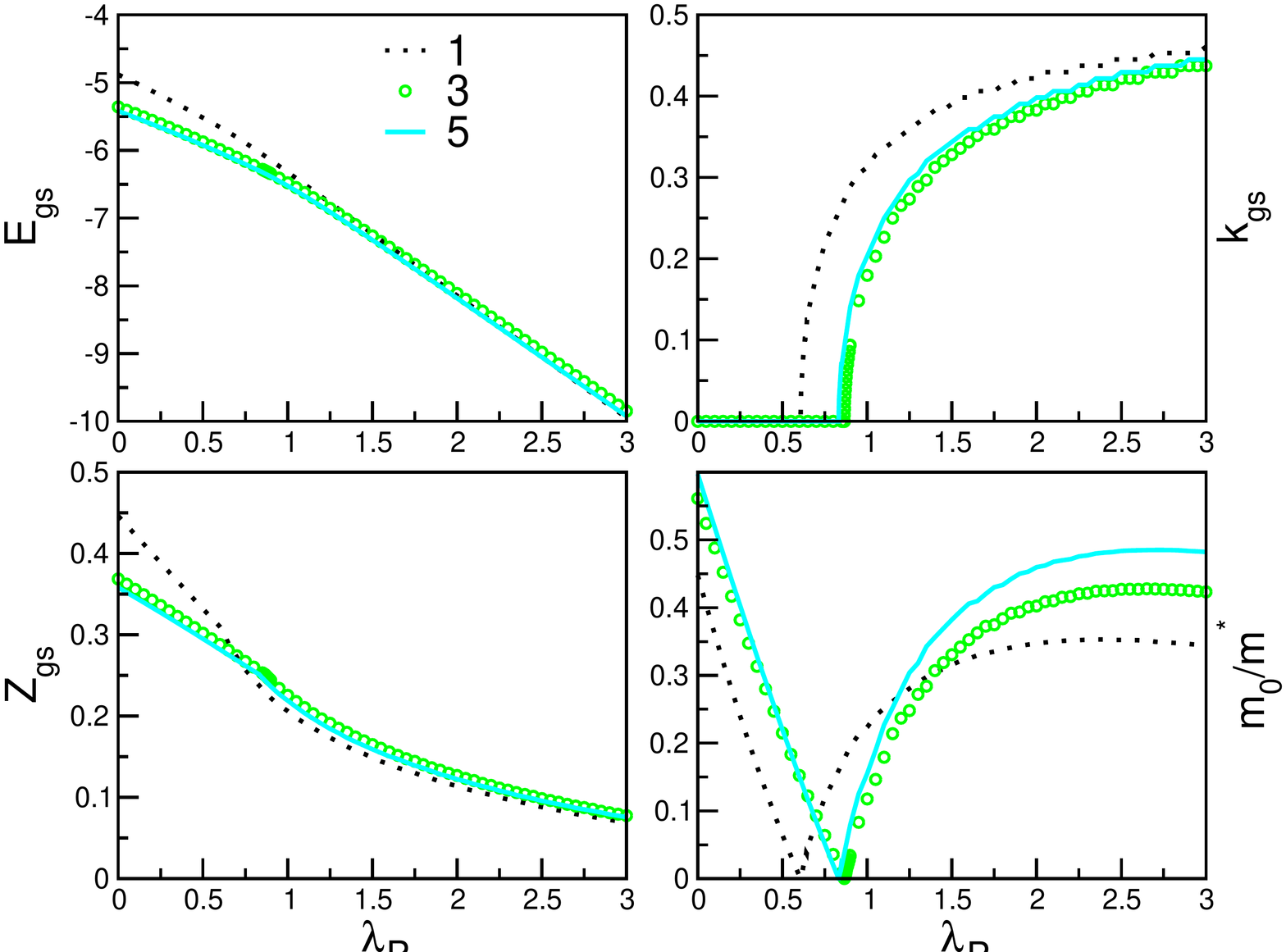}   
\vskip 0.1 cm 
\caption{Polaron ground state properties. The four panels show the ground-state energy $E(k_{gs})$, GS momentum $k_{gs}$, quasiparticle weight $Z_{gs}$ and inverse effective mass change {\em vs.} Peierls coupling $\lambda_P$, at a fixed and large $\lambda_h=2$ and for a large $\Omega=3$. The ranges of the EHM are as follows: 1 (black dotted line), 3 (green filled circles) and 5 (solidy cyan lines).   
  \label{fig7}}
\end{figure}

However, the interplay between the effects of the extended range and those of a dual coupling are, in fact, even more intricate, as can be seen from Fig. \ref{fig7}. Here we consider how the  GS properties of the polaron, specifically its energy $E(k_{gs})$, GS momentum $k_{gs}$, quasiparticle weight $Z_{gs}$ and inverse effective mass change with increasing Peierls coupling $\lambda_P$, at a fixed but large $\lambda_h=2$ and for a large $\Omega=3$. For these values the polaron is closer to the small polaron regime (note that $Z_{gs} < 0.5$ at $\lambda_P=0$ for the Holstein model, label 1) and one would therefore anticipate a reduced sensitivity to range. The results show a significant change when coupling to the nn sites is turned on (label 3). This is reasonable, given that the Peierls coupling also involves phonons on the nn sites but with different signs, so 'interference' effects can be expected (we further comment on this below). The less expected fact is that  for larger $\lambda_p$ there is another sizable change when turning on the coupling to the 3rd nn sites (label 5), as most clearly seen in the value of the effective polaron  mass. We find this hard to explain in any simplistic way.

Figure \ref{fig14} provides a  different  way to analyze the sensitivity to the range of the coupling. Here we plot the critical value $\lambda^*_P$ at which $k_{gs}$ has the transition from 0 to a finite value, \cite{Dominic} for several EHM strenghts $\lambda_{eh}=\lambda_h$ (curves of different colors in each panel). The various panels correspond to various ranges of the EHM coupling.

The black dotted curves are for Peierls only coupling ($\lambda_h=0$), thus they are the same in all panels. They agree with the known result from Ref. \onlinecite{Dominic} and are provided as guides to the eye. The other three curves in each panel are for increasing values of the EHM coupling $\lambda_h$. For the dual Holstein+Peierls coupling (panel labelled 1), an increase of $\lambda_h$ results in a decrease of $\lambda^*_P$, in agreement with the results reported in Ref. \onlinecite{DominicPRB}. Extending the Holstein coupling to nn sites (panel labelled 3) has a drastic effect: now  $\lambda^*_p$ increases strongly with increasing $\lambda_h$ nearly everywhere in the parameter space. (The  exception is for the smaller value $\lambda_h=0.5$ for a narrow range of phonon frequencies close to $\Omega \approx 0.5$, where $\lambda^*_p$ is slightly below the value for the pure Peierls model.) Further extending the range to 2nd and to 3rd nn sites (panels labelled 5 and 7, respectively) has less severe impact. For small $\Omega$ there is a visible decrease of $\lambda^*_P$ with increasing range for the strongest $\lambda_h=2$, but everywhere else the results for $\lambda^*_P$ appear to be converged -- at least at this scale. This supports the view that the strongest sensitivity to the range of the coupling is observed as one moves towards the adiabatic limit. However, we emphasize again that even for parameters like those used in Fig. \ref{fig7}, which show little difference in the value of $\lambda^*_P$ between 2nd and 3rd nn ranges, the changes in $m^*$ are much more substantial.

\begin{figure}[t]
\centering \includegraphics[width=0.7\columnwidth,angle=-90]{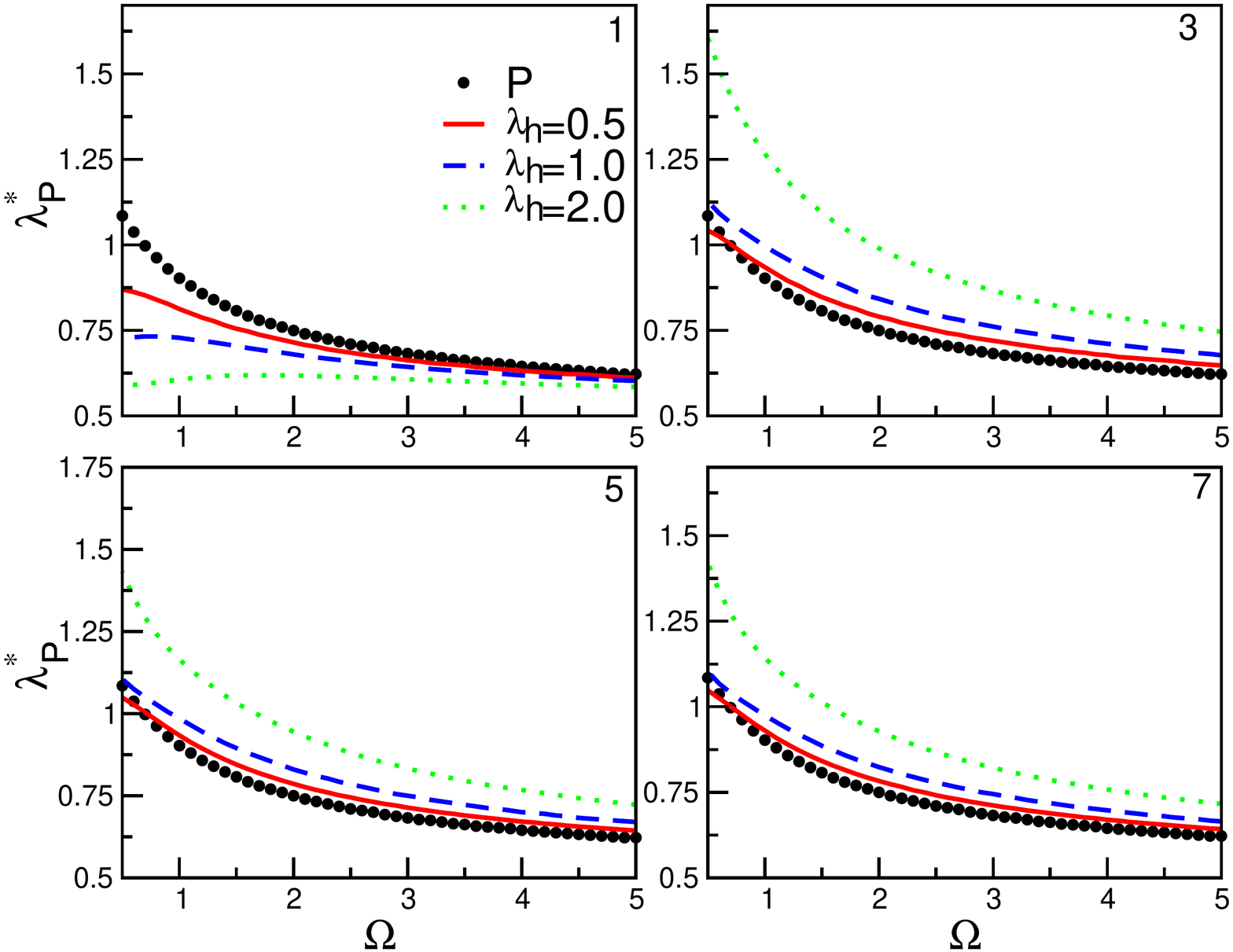}
\caption{The critical coupling $\lambda_{P}^{*}$ vs $\Omega$ for different ranges of the extended Holstein 
coupling: 1 (top left), 3 (top right), 5 (bottom left) and 7 (bottom right). The EHM is set to $\lambda_{eh}=0$ (black filled circles), $\lambda_h$ =$0.5$ (red solid lines),  $\lambda_h$ =$1$ 
(blue dashed lines) and  $\lambda_h$ =$2$ (green dotted lines).   
\label{fig14}}
\end{figure}

Our conclusion is that insofar as a dual EHM+P coupling is concerned, the polaron properties' sensitivity to the specific modeling of the extended coupling should certainly be a big concern in the adiabatic limit. As discussed, this is in line with expectations of a  more extended polaron cloud in this limit. However, our results show that the sensitivity to the range of the coupling could be significant even rather far from the adiabatic limit, and that it varies for different properties. Furthermore, even in parts of the parameter space where the results converge fast with increasing range, we find that adding nn coupling makes a significant difference and the results are very different from those due to on-site coupling only. 

\subsection{EBM+P results}

The significant differences observed everywhere when adding coupling to nn sites  motivates us to study how sensitive the results are to the particular  $g(q)$ coupling used. To do this, in this section we replace the EHM with EBM coupling. We remind the reader that EBM has zero on-site strength, so the range here starts from coupling to nn sites and can be extended to coupling to 2nd and 3rd nn sites (labels 3, 5 and 7, respectively). Also, we study three non-equivalent possible overall signs for these couplings, as illustrated in Fig. \ref{fig1}. 

In Figures \ref{fig15} and \ref{fig17} we show the shifted polaron dispersions energies $E_k-E_0$ vs. $k$ at $\Omega=3$ and $\Omega=0.5$, respectively. We remind the reader that  $\lambda_{ebm}=0.5$ means that the coupling $g_{bm}$ is chosen to be equal to the EHM $g$ corresponding to the same $\lambda_{eh}=\lambda_{ebm}$, ie the magnitude of the couplings to nn (and further sites, if allowed) remains the same as for the corresponding EHM. 

\begin{figure}[t]
\centering \includegraphics[width=0.7\columnwidth,angle=-90]{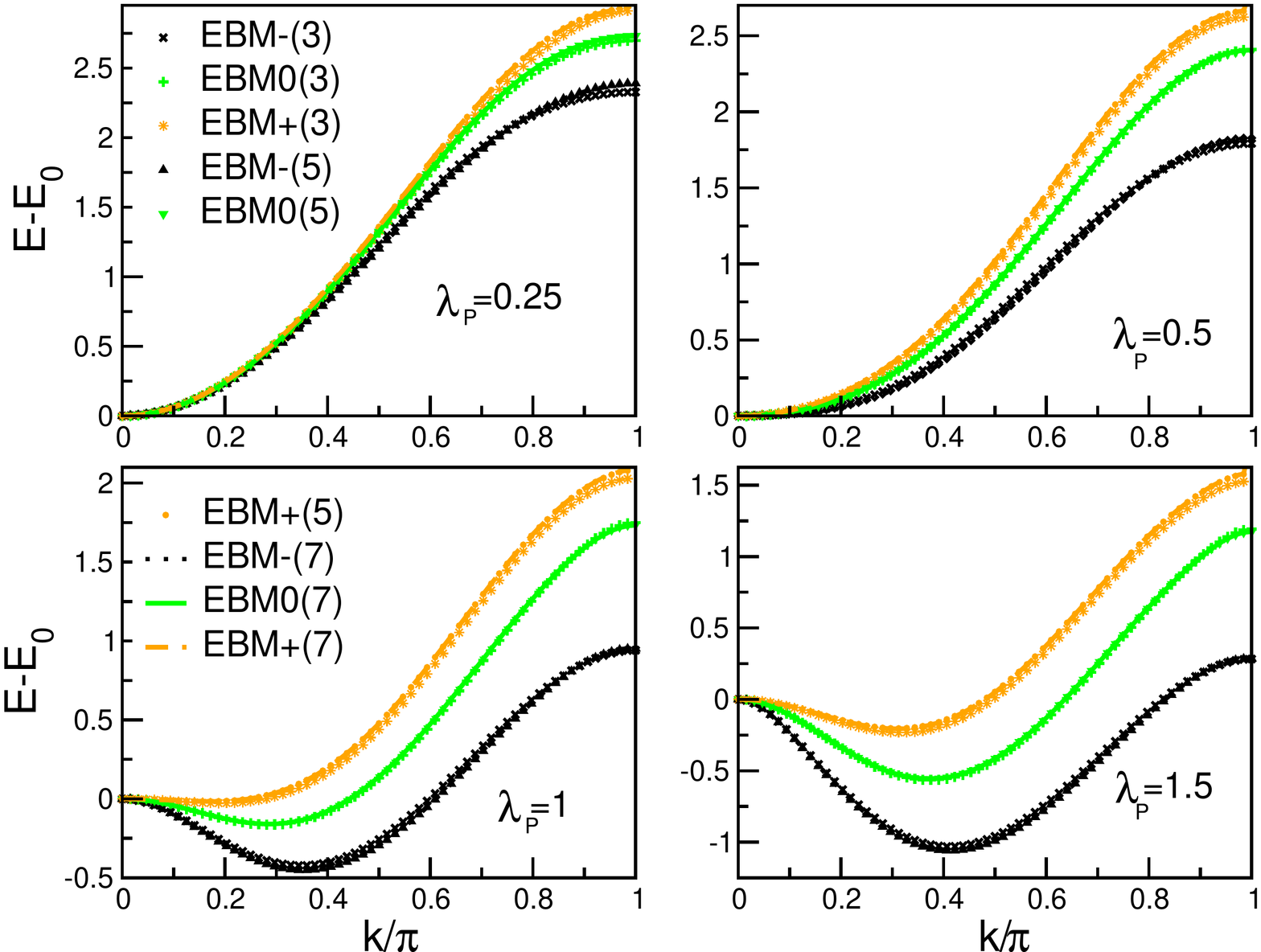}
\caption{Shifted polaron dispersion $E_k-E_0$ vs $k$ at $\Omega$=$3$ for $\lambda_{ebm}$=$0.5$ at four different $\lambda_{P}$ in the four panels, for the three EBM models $-,0,+$  (black, green, orange, respectively). The corresponding ranges (3,5, or 7) are indicated in the legend.
 \label{fig15}}
\end{figure}

\begin{figure}[b]
\centering \includegraphics[width=0.7\columnwidth,angle=-90]{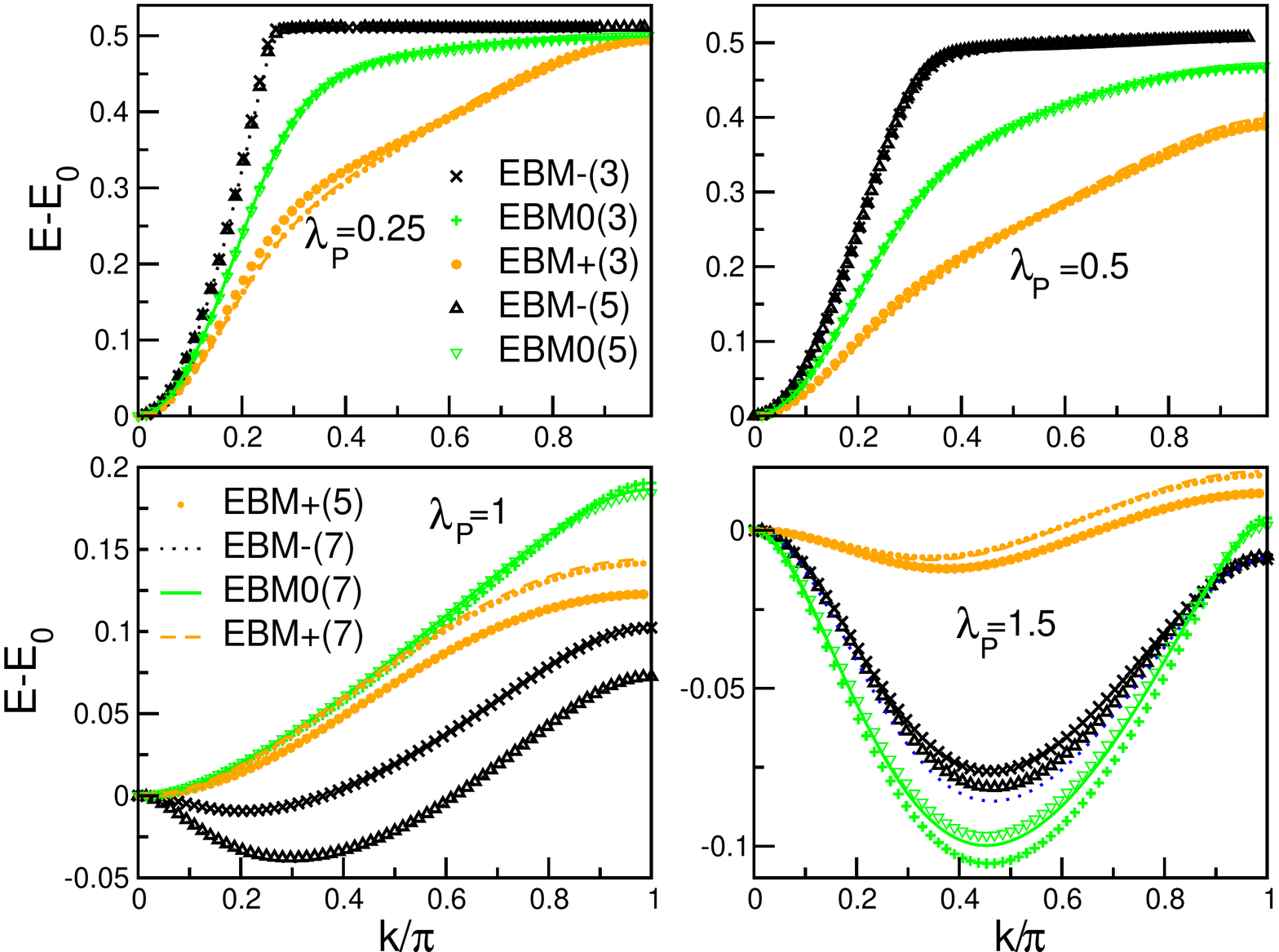}
\caption{Sames as Figure \ref{fig15} but for $\Omega=0.5$.
\label{fig17}}
\end{figure}

\begin{figure}[t]
\centering \includegraphics[width=0.7\columnwidth,angle=-90]{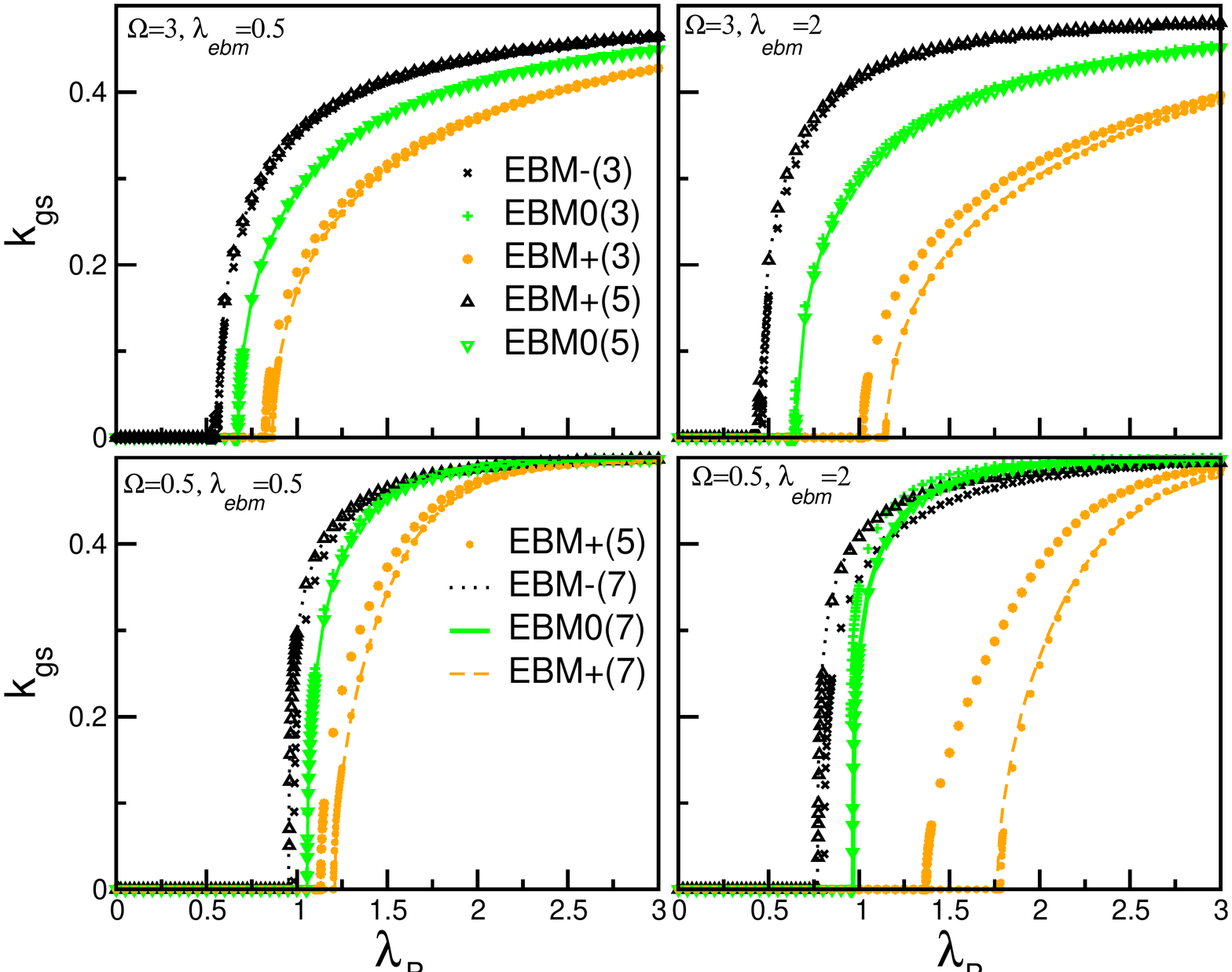}
\caption{Ground-state momentum $k_{gs}$ as a function of $\lambda_P$, when $\Omega=3$ and 0.5 (top panels and bottom panels, respectively) and for $\lambda_{ehm}=0.5$ and 2 (left panels and right panels, respectively). Results are for the three EBM models $-,0,+$  (black, green, orange, respectively). The corresponding ranges (3,5, or 7) are indicated in the legend.
\label{fig18}}
\end{figure}

\begin{figure}[b]
\centering \includegraphics[width=0.7\columnwidth,angle=-90]{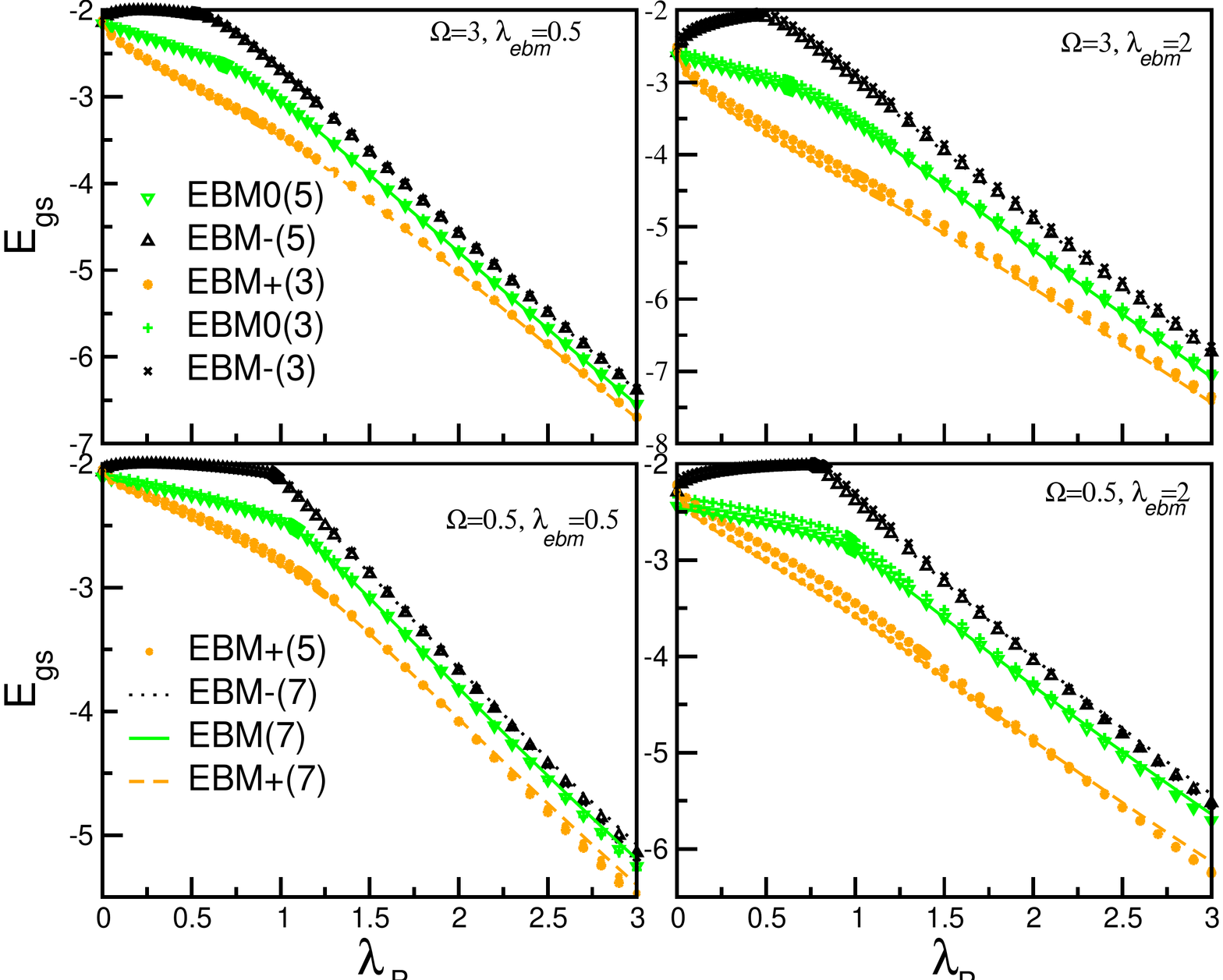}
\caption{Ground-state energy $E_{gs}$ as a function of $\lambda_P$. All parameters and symbols are as in Fig. \ref{fig18}.
\label{fig19}}
\end{figure}

\begin{figure}[t]
\centering \includegraphics[width=0.7\columnwidth,angle=-90]{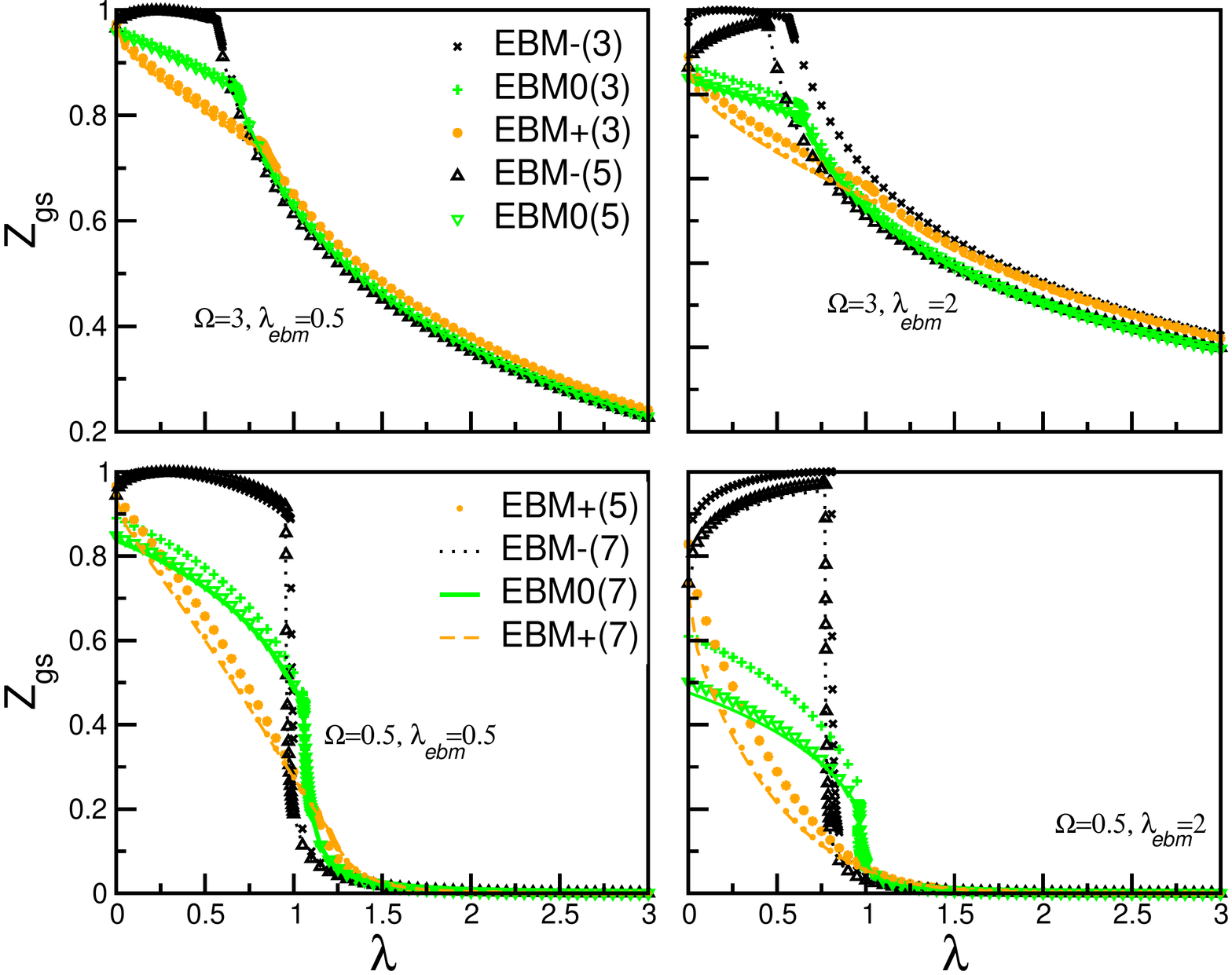}
\caption{Ground state quasiparticle weight $Z_{gs}$ as a function of $\lambda_P$. All parameters and symbols are as in Fig. \ref{fig18}.
\label{fig20}}
\end{figure}

\begin{figure}[b]
\centering \includegraphics[width=0.7\columnwidth,angle=-90]{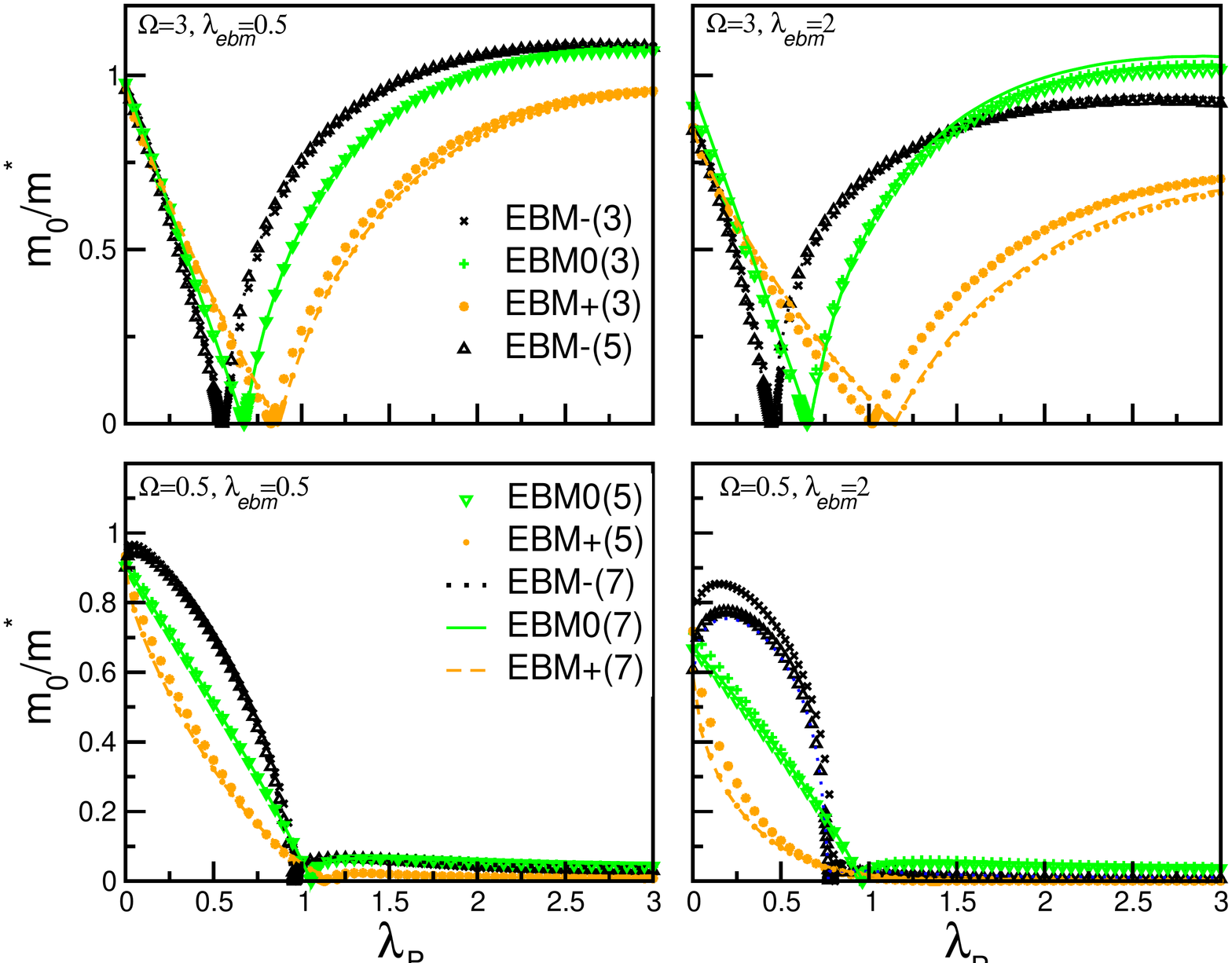}
\caption{Inverse effective polaron mass $m/m^{*} (k=k_{gs}$)  as a function of $\lambda_P$. All parameters and symbols are as in Fig. \ref{fig18}.
\label{fig21}}
\end{figure}

Before commenting on the sensitivity to the range of the coupling, it is interesting to note how peculiar are the shapes of some of these polaron dispersions, especially for the P+EBM+ model at weak and medium couplings in the adiabatic limit. The top left panel of Fig. \ref{fig17}, for instance, shows that the P+EBM- polaron dispersion has the usual expected shape at weak couplings, where the dispersion first roughly follows the bare dispersion and then flattens just below the polaron+one-phonon continuum that starts at $E_0+\Omega$. By contrast, the P+EBM+ polaron dispersion is located well below the continuum everywhere in the Brillouin zone, only reaching the continuum at $k=\pi$ for the weakest couplings considered. Given that the only difference is the overall sign of these EBM couplings, this  major difference is clearly a consequence of the interference between the BM and the Peierls couplings. we discuss this in more details below.

Similar to the results obtained with EHM+P, we see little sensitivity of the polaron dispersion to the range for the larger $\Omega$, whereas as the adiabatic limit is approached this sensitivity appears again, especially at stronger $\lambda_P$. What is very striking, though, are the significant differences between the results for the three different EBM couplings (also from those for the corresponding EHM+P model). This is further demonstrated in Figs. \ref{fig18}-\ref{fig21}, where we show the various GS properties of the polarons as a function of $\lambda_P$. All of these are quantitatively quite different for the three EBM flavors. They also show a different level of sensitivity to the range of the coupling, which depends not only on the 'flavor' of EBM but also on the ground-state property considered. For instance, in Fig. \ref{fig18}, the dual P+EBM+ model shows significant differences in all panels for going from BM coupling to only nn sites, to EBM coupling to 2nd nn, and also to 3rd nn sites. The dual P+EBM- model only shows this sensitivity as the adiabatic limit is approached when $\lambda_{ebm}$ is increased, while the P+EBM0 coupling is the least sensitive to range. On the other hand, the top right panel of Fig. \ref{fig20} reveals that for those parameters, the P+EBM- model has the larger sensitivity to range.

It is also interesting to note the unusual behavior of the P+EBM- model in the weak-coupling limit of $\lambda_P$, for the stronger $\lambda_{ebm}=2$ values. Here all ground-state quantities have rather unusual behavior, most strikingly $E_{gs}$ {\em increases} with $\lambda_P$ below the sharp transition. This is suggestive of a partial reciprocal cancellation of the effects of the Peierls and EBM+ couplings, consistent with previous work showing strong interference effects between the (short-range) BM and Peierls models.\cite{Roman} In fact, we can follow the arguments in Refs. \onlinecite{Dominic,Roman} to understand the behavior of these dual models in the strongly anti-adiabatic limit where $\Omega \gg \alpha, g_{bm}$. After projecting out the higher energy manifolds with one or more phonons, the resulting effective polaron dispersion is $E_k = \epsilon_p - 2t^* \cos k - 2t_2 \cos 2k$. For all three flavors of EBM, we find the expected $\epsilon_P = - {4 \alpha^2\over \Omega} - {g_{bm}^2\over \Omega}\sum_{|j|<M}^{} f^2_j$ and the phonon-mediated 2nd nn effective hopping $t_2= - \alpha^2/\Omega$ which is the driver of the sharp transition in the polaron GS properties.\cite{Dominic,DominicPRB} However, we also find a renormalization of the nn hopping.\cite{Roman} Specifically, $t^*=t$ for the P+EBM0 model, whereas $t^* = t \pm{2\alpha g_{bm}f_1\over \Omega}$ for the P+EBM$\pm$ couplings, demonstrating the interference between the Peierls and the nn BM couplings. For small $\alpha$ (weak $\lambda_P$) this renormalization of $t^*$ is the dominant change, explaining the already mentioned increase of $E_{gs}$ with $\lambda_P$ for the P+EBM- model. Of course, at larger $\lambda_P$, the cuadratic terms become more important, the dispersion becomes dominated by $t_2$ and the system transitions into the strong-coupling polaron regime. This argument also qualitatively explains the considerable differences in the values of $\lambda_P^*$ shown in Fig. \ref{fig22}. On the other hand, the sensitivity to the range of the EBM coupling, still clearly seen for the P+EBM+ model even at $\Omega=5$, is a higher-order effect whose explanation requires a more accurate approximation.

\begin{figure}[]
\centering \includegraphics[width=0.7\columnwidth,angle=-90]{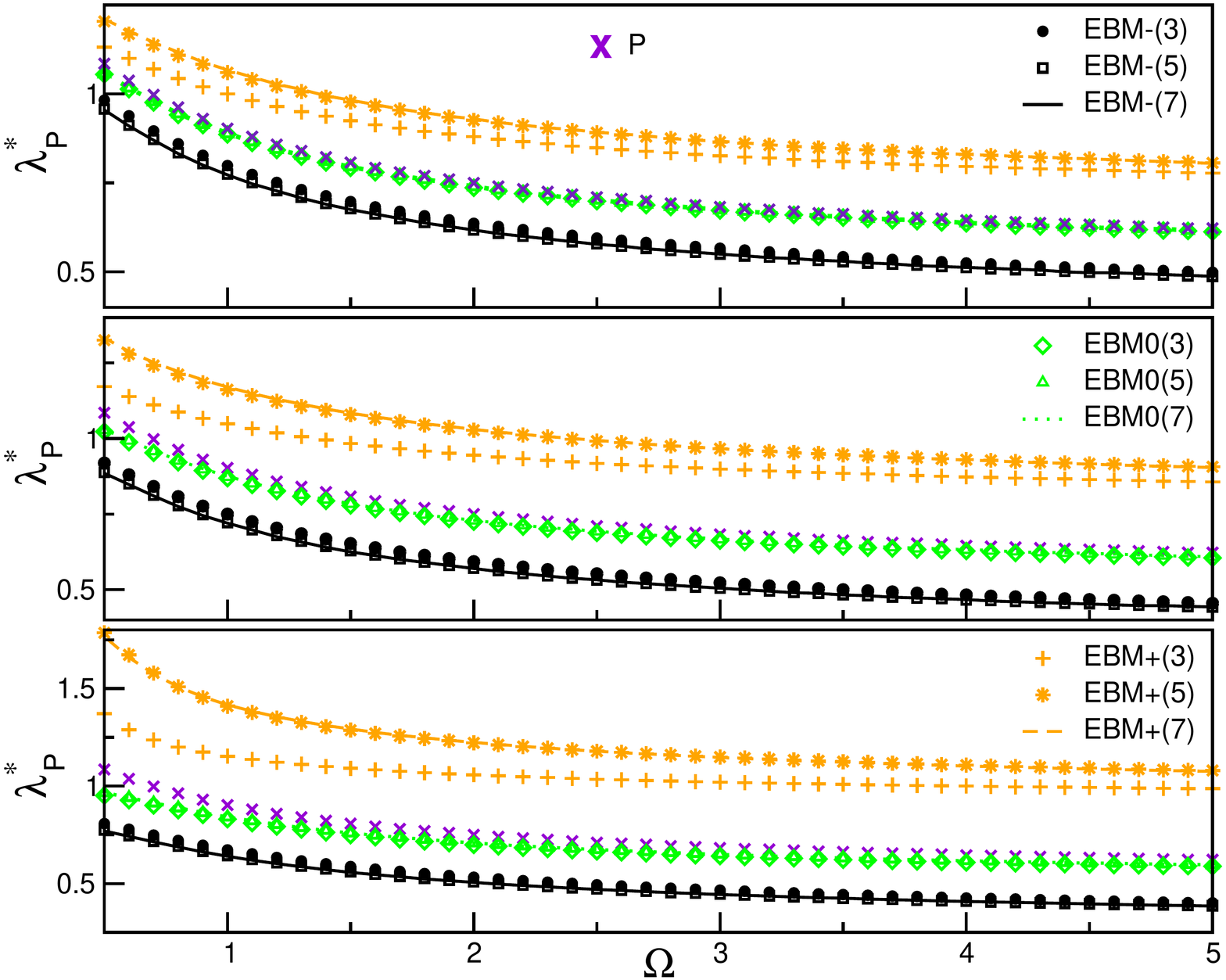} 
\caption{The critical coupling $\lambda_{P}^{*}$ vs $\Omega$, for $\lambda_{ebm}$=$0.5, 1$ and 2 in the top, middle and bottom panels. The violet cross symbols show the result for the pure Peierls case. All other symbols are the same as in Fig. \ref{fig18}.
\label{fig22}}
\end{figure}

\section{Summary and conclusions}

To the best of our knowledge, this work reports  the first   study of single polaron properties in dual models with Peierls + longer-range $g(q)$ couplings. For the latter, we considered both the extended Holstein coupling and three flavors of the extended breathing-mode coupling. In all cases, we find that the polaron dispersion and other physical properties like the polaron effective mass are quite sensitive to the spatial range of the extended coupling, especially as the adiabatic regime is approached. This is to be expected to some extent, given that the size of the polaron cloud increases as $\Omega/t$ decreases, but the persistence of this sensitivity even far from the adiabatic regime, for some quantities, is surprising. Furthermore, we find that the different extended $g(q)$ couplings can lead to very different results, even if they have equal magnitudes. These differences come from the different signs and different inversion symmetry (even or odd) of these couplings.  Such differences are expected because the nn BM coupling will interfere differently with the Peierls coupling (which is also sensitive to nn phonons) for the different signs and symmetries, as demonstrated in the anti-adiabatic expression for the polaron dispersion discussed above. Again, though, the surprising sensitivity to the range of the EBM coupling points to the generation of even longer-range effective hoppings, whose effect can be significant even at rather large $\Omega$ values. 

Taken together, these results point to the need to carefully consider the actual range and the actual form of the electron-phonon coupling, if a quantitative comparison with experiments is desired. While our work has been confined to the single-polaron limit, such significant differences between various longer-range models should be expected at finite carrier concentrations as well, although it is a matter of further research to understand how significant they are. We believe that doing this work is timely, in light of the recent studies\cite{ZX,Yao} on cuprate chains.

\acknowledgements M.C. appreciates access to the computing facilities of the DST-FIST (phase-II) project installed in the Department of Physics, IIT Kharagpur, India.  M. B. acknowledges support from the Natural Sciences and Engineering Research Council of Canada (NSERC), the Stewart
Blusson Quantum Matter Institute (SBQMI) and the
Max-Planck-UBC-UTokyo Center for Quantum Materials.

\end{document}